\documentclass[%
 reprint,
 amsmath,amssymb,
 aps,
prb,
]{revtex4-2}

\usepackage{graphicx}
\usepackage{dcolumn}
\usepackage{bm}
\usepackage{appendix}

\usepackage{color}
\usepackage{braket}
\usepackage{multirow}

\begin{document}

\preprint{APS/123-QED}

\title{Classical transport theory for the planar Hall effect with threefold symmetry}

\author{Akiyoshi Yamada}
\author{Yuki Fuseya}%
\affiliation{%
Department of Physics, Kobe University, Kobe 657-8501, Japan
}%




\date{\today}

\begin{abstract}
In recent years, the planar Hall effect (PHE) has become a key probe of Berry curvature and the anomalous Hall effect (AHE). Threefold-symmetric signals under in-plane fields are often attributed to such quantum mechanisms.
Here, we establish a purely classical origin for a three-fold-symmetric PHE.
The idea is simple yet decisive: a third-order expansion of the Boltzmann equation in the magnetic field reveals that the threefold component originates from the relative positions of the mirror planes in the crystals with respect to the measurement setups.
Remarkably, the threefold contribution should be ubiquitous because this symmetry condition can be realized across a broad range of crystals. 
Numerical estimates based on concrete models further show that its amplitude is comparable to that expected from the AHE.
\end{abstract}

\maketitle

The planar Hall effect (PHE)---Hall effect in the in-plane magnetic field---was first discovered by Goldberg theoretically~\cite{Goldberg1954}.
A characteristic angular dependence:
\begin{eqnarray}
\rho_{xy}(B,\phi)=\left[\rho_\perp(B)-\rho_\parallel(B)\right]\sin2\phi,
\end{eqnarray}
where $\phi$ is the azimuthal angle, was extracted by applying the Jones–Zener expansion~\cite{Jones-Zener} to the Boltzmann equation~\cite{Seitz1950} to the {\it second-order} term with the magnetic field. 
The following magneto-transport experiments on semiconductors well established this effect~\cite{Shibuya1954,1957shogenji}.
Nevertheless, recent literature occasionally claims that the contributions of the Lorentz force are absent in resistivity under the PHE setup, which is erroneous.
As Goldberg presented, this effect originates from the difference between longitudinal and transverse magnetoresistances, which is nothing but a consequence of the Lorentz force.

A modern view of the PHE has attracted significant attention in fields at the intersection of solid-state physics and particle physics. 
Electronic structures equivalent to the Weyl Hamiltonian are realized due to strong spin–orbit coupling in certain materials~\cite{Lu2015}. 
The lowest-energy states of such electron systems split into different chiralities under an external magnetic field, and an electric field parallel to the magnetic field selectively excites chirality-polarized carriers~\cite{Nielsen1983}. 
This phenomenon is the solid-state analogue of the chiral anomaly (CA), a breakdown of CP symmetry in particle physics, which highlights the fact that the anomaly can be probed through galvanomagnetic effects in condensed matter systems~\cite{Ong2021}.
The PHE was theoretically established as a probe of the anomaly~\cite{Nandy2017,Burkov2017}, prompting numerous subsequent experiments~\cite{Xiong2015,Kumar2018,Liang2018,Li2019,Yang2020PHE}.
However, a fundamental difficulty arises: the PHE originating from the chiral anomaly exhibits the same $\sin2\phi$ angular dependence as the PHE from the classical origin, making a strict distinction difficult.
Indeed, several observed PHE amplitudes can be quantitatively explained within the classical framework alone~\cite{Li2020,Yamada2021}.

In recent years, more and more extensive theoretical and experimental efforts have sought to extract peculiar information from PHE, especially topological features based on Berry curvature, and the PHE induced by the anomalous Hall effect (AHE) in magnetic materials, quantum AHE, or orbital magnetization has attracted particular interest~\cite{2013_Liu,2016_Ren,2020_Zyuzin,2021_Battilomo,2021_Cullen,2022_Li,2022_Sun,2023_Cao,2024_Wang}.
These significant responses include a threefold symmetric component in the angular dependencies:
\begin{eqnarray}
\rho^{(3)}_{xy}(B,\phi)=\rho_{3}(B)\sin3\phi.
\end{eqnarray}
Very recently, advances in high-quality thin-film growth and accurate voltage measurements under magnetic fields have enabled the experimental extraction of threefold components in Weyl-, Dirac-type electronic systems~\cite{nakamura2024,nishihaya2025anomaloushalleffectdirac,nishihaya2025spontaneousinplaneanomaloushall}, where pronounced Berry-phase effects are expected~\cite{2013_Liu,2016_Ren,2020_Zyuzin,2021_Battilomo,2021_Cullen,2022_Li,2022_Sun,2023_Cao,2024_Wang}.
Among these materials, semimetallic systems show even greater amplitude of the threefold angular dependence ($\sim \text{m}\Omega\text{cm}$) than that in metallic ones ($\sim\mu\Omega\text{cm}$).

At this point, two pressing questions arise. 
Is it impossible to find the threefold component in PHE from the classical point of view?
And why do some materials (such as Cd$_3$As$_2$~\cite{nishihaya2025anomaloushalleffectdirac} and ZrTe$_5$~\cite{Liang2018_AHE,2025_wang}) exhibit asymmetric components in PHE even without spontaneous time-reversal symmetry breaking and misalignment?

In this work, we construct a classical theory of the asymmetric components in the PHE, with particular focus on the threefold-symmetric contribution, which has been overlooked. 
Our central insight is remarkably simple: a {\it third-order} expansion of the Boltzmann equation in $B$ establishes a general framework for the threefold-symmetric PHE.
This expansion explicitly reveals the conditions under which the threefold-symmetric PHE can emerge from the classical origin governed by the Lorentz force. 
Strikingly, its presence is governed not by rotational symmetry but by mirror symmetry with respect to planes defined by the current direction.
We confirmed this point by calculating the threefold component with the effective mass model and the tight-binding model.
Going further, by incorporating not only symmetry but also detailed structural features such as anisotropy and carrier density, we predict the amplitude of the classical threefold PHE. 
The amplitude in anisotropic metals, semimetals, and semimetals can reach values comparable to those previously attributed to an AHE origin.

\begin{table}[b]
\caption{\label{table_parity_cond} The numbers of differentiation operation in the $\Lambda_{ij}$ and the parity of the $\Lambda_{ij}$ along $k_x, k_y, k_z$ directions in setups in Fig.~\ref{fig_mirrors}.}
\begin{center}
\begin{tabular}{c|ccc|ccc|ccc}
    \hline
    & & $\partial^l_{k_x}\partial^m_{k_y}\partial^n_{k_z}$ & & & setup1 &  & & setup2&\\[2pt]\hline
    & $l$ & $m$ & $n$ 
      & $ k_x$ & $k_y$ & $ k_z$
      & $k_x$ & $ k_y$ & $ k_z$\\[2pt] \hline\hline 
     $\Lambda_{xx}$ & 1 & 4 & 3
      & odd & even & odd
      & odd &  e $+$ o & e $+$ o\\[2pt]
     $\Lambda_{yy}$ & 4 & 1 & 3
     & even & odd & odd
     & {\bf even} & {\bf e} $+$ o & {\bf e} $+$ o\\[2pt]
     $\Lambda^{(1)}_{xy}$,$\Lambda^{(2)}_{xy}$ & 2 & 3 & 3
     & even & odd & odd
     & {\bf even} & {\bf e} $+$ o & {\bf e} $+$ o\\[2pt]
     $\Lambda^{(1)}_{yx}$,$\Lambda^{(2)}_{yx}$ & 3 & 2 & 3
     & odd & even & odd
     & odd &  e $+$ o & e $+$ o\\[2pt]
    \hline
\end{tabular}
\end{center}
\end{table}

First, we demonstrate how a classical mechanism governed by the Lorentz force can give rise to a threefold-symmetric PHE. 
Jones and Zener provided a solution for the Boltzmann equation that can be systematically expanded to arbitrary order in the field~\cite{Jones-Zener}.
We employ this formalism to calculate the Hall conductivity. 
The solution can be written as follows:
\begin{eqnarray}
\sigma_{yx}&=&-\frac{2e^2\tau}{(2\pi)^3}\int dk\,v_y(1+\hat\Omega)^{-1}\left\{{v_x} \frac{\partial f}{\partial \varepsilon}\right\}\label{eq_sigma_full}\\
\hat\Omega&=&\frac{e\tau}{\hbar}\left\{(\bm v\times{\bm B})\cdot\nabla_k\right\},
\end{eqnarray}
where ${\bm v},f$ are the group velocity ($v_{i}\equiv\hbar^{-1}\partial\varepsilon/\partial k_{i}$) and the Fermi distribution function.
$e>0$, $\hbar$, and $\tau$ are the elemental charge, the reduced Planck constant, and the carrier relaxation time, respectively. 
Seitz and Goldberg obtained the twofold-symmetric PHE by truncating the expansion at second order~\cite{Seitz1950,Goldberg1954}, $O(B^2)$, but, to our knowledge, no expansion beyond this order had previously been attempted.
To assess the presence of a threefold-symmetric component under a rotating in-plane field, we assume an in-plane magnetic field $(B\cos\phi,B\sin\phi,0)$ and expand the response to third order in $B$. This expansion yields four terms $\{\cos^3\phi,\,\sin^3\phi,\,\cos^2\phi\sin\phi,\,\cos\phi\sin^2\phi\}$, each of which reduces, via the triple-angle identities: $\sin3\phi=3\sin\phi-4\sin^3\phi,\,\cos3\phi=4\cos^3\phi-3\cos\phi$, to contributions proportional to $\sin 3\phi$ and $\cos 3\phi$. Hence, the cubic-in-field response to an in-plane magnetic field constitutes the threefold-symmetric component of the PHE.
The third-order expansion of Eq.~(\ref{eq_sigma_full}) is written as follows.
\begin{eqnarray}
\sigma_{yx}^{(3)}=-\frac{e^5\tau^4}{16\pi^3}&&\left[\braket{\Lambda_{xx}+\Lambda_{yx}}\cos3\phi+\braket{\Lambda_{yy}+\Lambda_{xy}}\sin3\phi\right]B^3,\nonumber\\\label{eq_C3comp}
\end{eqnarray}
\begin{eqnarray}
\Lambda_{xx}&=&K_{yz}\left[\alpha_{xy}v_z-\alpha_{xz}v_y\right]{v_y},\nonumber\\
\Lambda_{yy}&=&K_{xz}\left[\alpha_{xx}v_z-\alpha_{xz}v_x\right]{v_y},\nonumber\\
\Lambda_{xy}&=&\Lambda_{xy}^{(1)}+\Lambda_{xy}^{(2)},\quad \Lambda_{yx}=\Lambda_{yx}^{(1)}+\Lambda_{yx}^{(2)},\label{eq_lambda_func}\\
\Lambda_{xy}^{(1)}&=&-2(\alpha_{xz}\alpha_{yz}-\alpha_{xy}\alpha_{zz})\left[\alpha_{xz}v_y-\alpha_{xy}v_z\right]v_y,\nonumber\\
\Lambda_{yx}^{(1)}&=&-2(\alpha_{xz}\alpha_{yz}-\alpha_{xy}\alpha_{zz})\left[\alpha_{xz}v_x-\alpha_{xx}v_z\right]v_y,\nonumber\\
\Lambda_{xy}^{(2)}&=&K_{yz}\left[\alpha_{xz}v_x-\alpha_{xx}v_z\right]{v_y},\nonumber\\
\Lambda_{yx}^{(2)}&=&K_{xz}\left[\alpha_{xz}v_y-\alpha_{xy}v_z\right]{v_y},\quad K_{ij}\equiv \alpha_{ii}\alpha_{jj}-\alpha_{ij}^2,\nonumber
\end{eqnarray} 
where $\alpha_{ij}\equiv\hbar^{-2}\partial^2\varepsilon/\partial k_{i}\partial k_{j}$, the bracket $\braket{..}$ represents $\int dk\,..(df/d\varepsilon)$, and $K_{ij}$ is the Gaussian curvature.
For simplicity, we assume that higher-order derivatives of the effective mass become negligibly small. 
However, including such terms does not alter the subsequent arguments. 

\begin{figure}[t]
\begin{center}
\includegraphics[width=8.5cm]{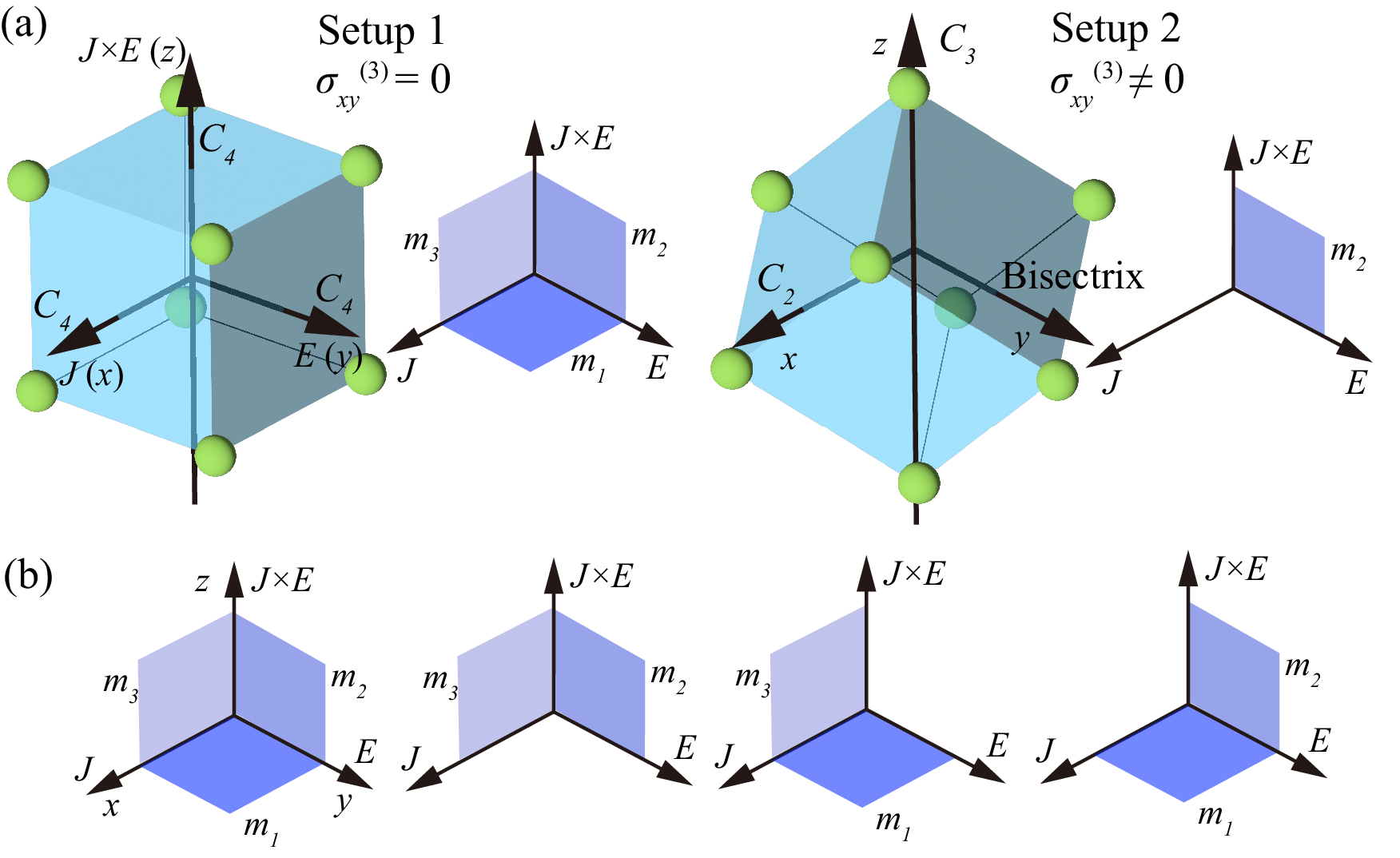}
\caption{\label{fig_mirrors} (a) Setups that show no $\sigma_{xy}^{(3)}$ (left) and finite $\sigma_{xy}^{(3)}$ (right) in cubic crystals. (b) Schematic images of sufficient condition where $\sigma_{xy}^{(3)}$ vanishes. $m_{1,2,3}$ denotes the mirror planes in the crystals, and $x,y,z$ axes corresponds to the directions of $\mathbf{j}$, $\mathbf{E}$, and $\mathbf{j}\times \mathbf{E}$ respectively. }
\end{center}
\end{figure}

Here we define the direction of the current~($\mathbf{j}$), the electric field~($\mathbf{E}$), and their normal~($\mathbf{j}\times\mathbf{E}$) as $x,y,$ and $z$ respectively.
The mirror symmetries with respect to planes associated with these axes ($yz$, $zx$, and $xy$) play a crucial role.
For the integral in Eq.~(\ref{eq_C3comp}) to be finite, the integrand $\Lambda_{ij}$ must contain an even component with respect to each of $x$, $y$, and $z$. 
$\Lambda_{ij}$'s parity is fixed by the parity of $\varepsilon(\bm{k})$ together with the differentiation orders.
Since differentiation with respect to $k$ flips the parity and each $\Lambda_{ij}$ contains odd-order derivatives along at least two directions~[Table~\ref{table_parity_cond} ($\partial_{k_x}^\ell\partial_{k_y}^m\partial_{k_z}^n$)], dispersion with even parity in those directions renders $\Lambda_{ij}$ odd ($\varepsilon$: $\text{even}\Rightarrow\Lambda_{ij}$: $\text{odd}$), and the integral vanishes. 
By contrast, a dispersion that breaks mirror symmetry contains both even and odd parts; differentiation does not remove this mixture ($\varepsilon$: $\text{even + odd}\Rightarrow\Lambda_{ij}$: $\text{even + odd}$), so $\Lambda_{ij}$ retains an even component and the integral remains finite.
We give explicit examples.
Consider a cubic crystal with all fourfold axes aligned with $x$, $y$, and $z$~[Fig.~\ref{fig_mirrors}(a) left]; odd-order differentiation of the dispersion alters parity, so every $\Lambda_{ij}$ is odd in at least two axes [Table~\ref{table_parity_cond} (setup1)]. 
The integration in Eq.~(\ref{eq_C3comp}) then vanishes, and no threefold component appears. 
Conversely, only a single mirror symmetry is conserved when the threefold axis is aligned with $z$~[Fig.~\ref{fig_mirrors} (a) right].
Several $\Lambda_{ij}$’s acquire even parts in all directions~[Table~\ref{table_parity_cond} (setup 2)], survive the integration, and produce a finite threefold contribution to the PHE.

Consequently, a threefold PHE becomes observable when mirror symmetry is broken with respect to two of the three planes defined by the current, the voltage, and the normal, whereas the presence of two or three mirror planes forces the threefold component to vanish identically [Fig.~\ref{fig_mirrors}(b)].

Next, we actually demonstrate the threefold PHE with a simple model.
A multi-ellipsoid model allows us to treat multiple Fermi surfaces~\cite{MacKey1969,Aubrey1971,Zhu2018,Mitani_2020}, their anisotropies, and the sign of the charge within the effective-mass approximation. 
The following expression gives the conductivity tensor for one Fermi surface:
\begin{eqnarray}
\hat{\sigma}_i&=&e n_i\left(\hat{\mu_i}^{-1}\pm\hat{B} \right)^{-1},\label{eq_conductivity_ellipsoid}
\end{eqnarray}
where the magnetic field is expressed by a tensor $B_{ij}=-\varepsilon_{ijk}B_k$ and $\varepsilon_{ijk}$ is the Levi-Civita symbol.
$\mu_{ij}=e\left(\tau/m^*\right)_{ij}$ is the mobility tensor, where $n_i$ and $m^*$ are the carrier density and effective mass, respectively. The sign $+$($-$) corresponds to the hole (electron) carriers.
The resistivity is obtained by taking the inverse of the total conductivity tensor: $\hat{\rho}= \left(\sum_i \hat{\sigma}_i \right)^{-1}$.
The anisotropies and symmetries of the Fermi surfaces are reflected in the calculation by the arrangement of the mobility tensor.
This method is non-perturbative with respect to the field, hence the threefold components demonstrated here remain even considering up to $O(B^\infty)$.

\begin{figure}[t]
\begin{center}
\includegraphics[width=7.6cm]{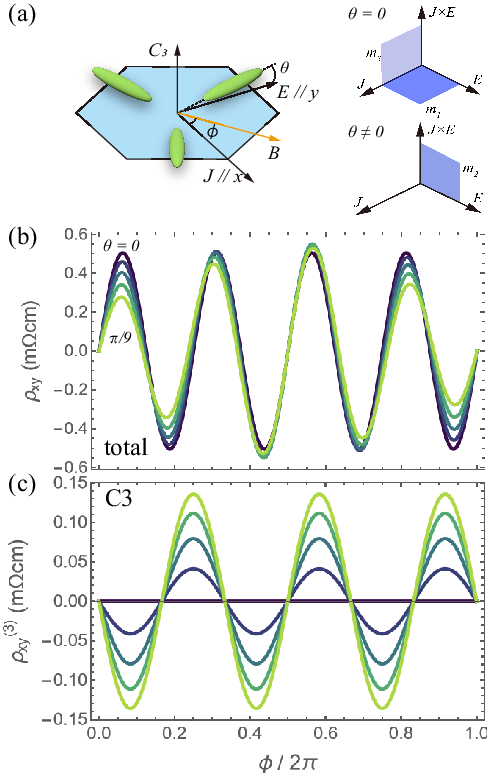}
\caption{\label{fig_qual} (a) Schematic images for ellipsoidal Fermi surfaces distributed around the $C_3$ axis~(left). The mirror planes with $\theta=0$ and $\theta\neq0$~(right). (b) Calculated angular dependence of PHE with several values of $\theta$. (c) Extracted the three-fold component from the PHE. $n_{1,2,3}=10^{17}$ cm$^{-3}$, $B=0.3$ T, $\mu_0=14$ T$^{-1}$, and $\beta=0.005$.}
\end{center}
\end{figure}

We consider a metal or degenerated semiconductor with electron carriers and arrange anisotropic ellipsoidal Fermi surfaces in a threefold-symmetric configuration [Fig.~\ref{fig_qual}(a)]. 
The mobility tensor of each surface is constructed by rotating an anisotropic mobility represented by a diagonal tensor: $(\mu_{11},\mu_{22},\mu_{33})=(\mu_0,\beta\mu_0,\mu_0)$ by $0^\circ$, $120^\circ$, and $240^\circ$ within the $xy$ plane.
We introduce the out-of-plane tilt angle $\theta$ of the Fermi surfaces relative to the $xy$ plane. 
Rotation of the magnetic field within the $xy$ plane yields the angular response shown in Fig.~\ref{fig_qual}(b). 
As $\theta$ increases from $0$, intricate components are included in the angular dependence. 
Extracting the threefold component shows that it accounts for roughly $25\%$ of the maximum amplitude [Fig.~\ref{fig_qual}(c)]. 
A configuration in which the ellipsoidal Fermi surfaces lie strictly in the $xy$ plane ($\theta=0$) exhibits a clear fourfold symmetry in the PHE. 
The preceding analysis indicates that the threefold component is governed not by rotational symmetry but by the presence or absence of mirror symmetry, as shown in Eq.~(\ref{eq_C3comp}); a threefold arrangement of Fermi surfaces alone does not generate a threefold-symmetric PHE. 
Tilted Fermi surfaces out of the rotation plane, which is equivalent to a cubic crystal oriented with its $C_3$ axis (body diagonal) along $z$, reduces the number of mirror planes to a single one and thereby enables a finite threefold contribution. 
Furthermore, the simple effective-mass model gives this contribution analytically, and one can see the contribution remains non-zero only for finite $\theta$ (see Supplemental Information).

\begin{figure}[b]
\begin{center}
    \includegraphics[width=8cm]{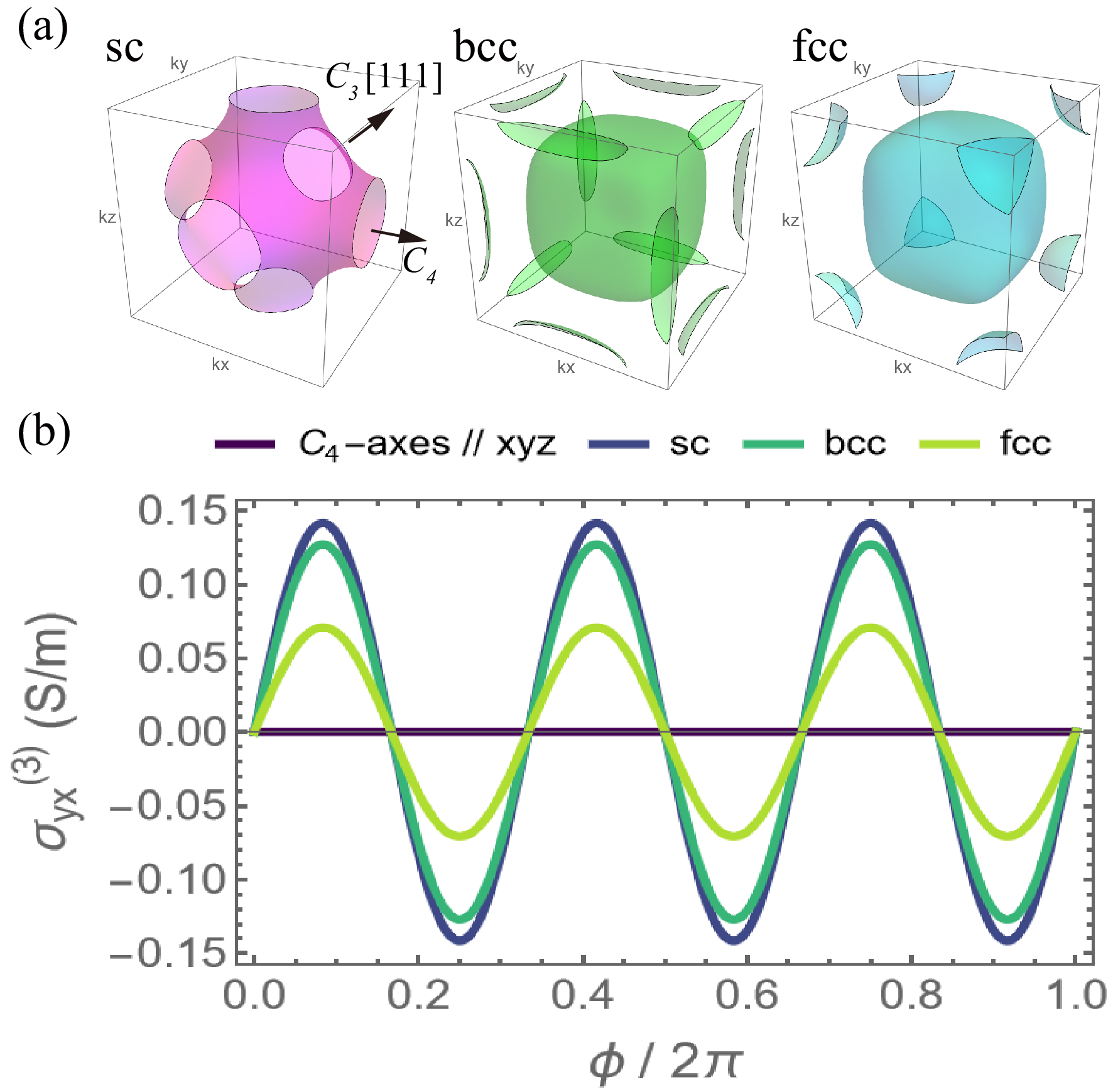}
\caption{\label{fig4_TB_Oh} (a) Fermi surface in a sc, fcc, and bcc lattice. (b) Three-fold angular dependence of $\sigma_{yx}$ in cubic systems. $E_F/t=-0.5$ and $k_BT/t=2.8\times10^{-2}$ where $t$ is the hopping parameter set to $30$ meV. The carrier lifetime $\tau$ is $1$ ps.}
\end{center}
\end{figure}

The point is alignment: the crystal–probe orientation dictates the threefold component, with applicability across a wide class of materials.
This principle becomes transparent when we measure the PHE in a cubic crystal with cubic symmetry in two distinct configurations. 
We consider $\mathrm{sc}$, $\mathrm{fcc}$, and $\mathrm{bcc}$. 
Figure \ref{fig4_TB_Oh}~(a) shows the Fermi surfaces obtained from the tight-binding model with nearest-neighbor hopping. 
We note that the following discussion remains fully general upon including further-neighbor hoppings or additional orbitals. 
The cubic symmetry hosts three $C_4$ axes perpendicular to mirror planes and the $C_3$ axes along the body diagonals.
We compare the PHE when the $C_4$ axes are chosen parallel to the $\mathbf{j}$~$(x)$, $\mathbf{E}$~$(y)$, and $\mathbf{j}\times\mathbf{E}$~$(z)$ axes with the case in which the $C_3$ axis is aligned with $z$ and the $C_2$ axes are aligned with the $x$ axes.

The threefold component computed according to Eq.~(\ref{eq_C3comp}) is shown in Fig.~\ref{fig4_TB_Oh}~(b). 
When the $C_4$ axes are set along the $x y z$ axes, $\sin3\phi$ dependence vanishes due to the parity profiles of the energy as described in Eq.~(\ref{eq_C3comp}). 
In contrast, when we take the $C_3$ axis along $\mathbf{j}\times \mathbf{E}$~(z) direction, the component remains finite. 
It renders this situation qualitatively equivalent to the threefold ellipsoidal arrangement with $\theta\neq0$. 
With this in mind, we can predict that the emergence of threefold anisotropy in the PHE is far from exceptional. 
Rather, it constitutes a ubiquitous response that naturally arises from the breaking of mirror symmetries—the common structural feature in crystals.

\begin{figure}[t]
\begin{center}
    \includegraphics[width=7.6cm]{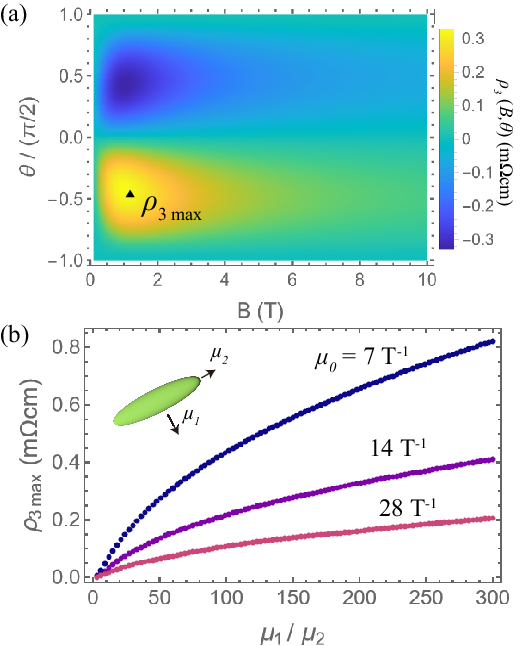}
\caption{\label{fig_quant} (a) Color map of the amplitude of the three-fold PHE in the three-ellipsoidal model as a function of the magnetic field and tilt angle. The carrir density is $3\times10^{17}$ cm$^{-3}$. $\mu_0=14$ T$^{-1}$ and $\beta=\mu_1/\mu_2=0.005$. (b) Maximum values of the three-fold PHE as a function of anisotropy in the mobility.}
\end{center}
\end{figure}

Finally, we discuss the amplitude of the threefold-symmetric component.
Is the classically derived PHE experimentally observable? 
In short, the answer is ``yes'', and one determinant is the anisotropy of the mobility.
Figure~\ref{fig_quant}~(a) maps the threefold component in the three-ellipsoidal model as a function of field and $\theta$. 
It exhibits a maximum at specific values of the field and the tilt, and this maximum increases with the mobility anisotropy up to the order of m$\Omega$cm~[Fig.~\ref{fig_quant}(b)]. 
In the high-field limit, the threefold component decays. 
Another determinant is the coexistence of charge in the carrier.
Semimetals hosting both electrons and holes show a similar trend in $B-\theta$ mapping; however, the field at which the maximum occurs depends sensitively on the density of excess charge. 
In clean samples with small deviation from the perfect compensation, the threefold component can persist beyond the available fields and attains a magnitude up to several hundreds of m$\Omega$cm, which is larger than in the single-carrier case (See {\it Supplemental Information}). 
In both semiconductors and semimetals, the amplitude of the threefold component lies well within experimental reach and, remarkably, is comparable to values previously attributed to the anomalous Hall effect.

We conclude that the threefold-symmetric PHE component has a purely classical origin—the breaking of multiple mirror symmetries exposed by a third-order expansion of the linearized Boltzmann equation.
Crucially, this condition requires neither spontaneous time-reversal-symmetry breaking nor Berry phase, but is instead determined purely by the crystalline symmetry and the experimental setup. 
All trigonal, triclinic, and monoclinic systems satisfy this condition, ensuring that the threefold-symmetric PHE cannot be eliminated regardless of the measurement geometry. 
It is particularly intriguing that all crystals in which threefold symmetry has been experimentally observed in the PHE also possess intrinsic threefold symmetry within the plane of magnetic-field rotation~\cite{nakamura2024,nishihaya2025anomaloushalleffectdirac,nishihaya2025spontaneousinplaneanomaloushall}.

In EuCd$_2$As$_2$, Cd$_3$As$_2$, and SrRuO$_3$, PHE measurements employ Hall bars created in the plane perpendicular to the crystallographic threefold (or approximate threefold) axis~\cite{nakamura2024,nishihaya2025anomaloushalleffectdirac,nishihaya2025spontaneousinplaneanomaloushall}, and the field–angle dependence relative to the voltage direction follows $\cos 3\varphi$, which corresponds to $\sin3\phi$ with the angle measured from the current direction ($\varphi=\phi+\pi/2$).
Under this configuration, the crystal retains mirror symmetry only with respect to the plane perpendicular to the current, corresponding to the case in Fig.~\ref{fig_mirrors} where only the $m_2$ remains. 
Consequently, the angle dependence of the PHE inferred from Table~\ref{table_parity_cond} is $\sin 3\phi$ in agreement with experiment.

We have also provided quantitative estimates of the threefold component for both semiconductors and semimetals. 
In both cases, the predicted amplitudes are sufficiently large to be experimentally observable and are consistent with previously measured resistivity values ($\sim\mu\Omega\text{cm}$ for metals and $\sim m\Omega\text{cm}$ for semimetals). 
Moreover, we have elucidated why semimetals host substantially larger threefold components than semiconductors: in semiconductors, the amplitude increases gradually with carrier anisotropy, and in semimetals, it exhibits further enhancement.

We hope that this work not only provides a fresh perspective on magnetotransport in magnetic fields but also enriches the interpretation of existing observations, thereby contributing new insights to the broader field of emergent quantum materials.

We thank M. Tokunaga, H. Sakai, H. Kotegawa, M. Uchida, A. Nakamura, Y. Deguchi, and A. Nevidomskyy for helpful comments and discussions.
This work is supported by JSPS KAKENHI (Grant No. 23H04862, No. 23H00268, No. 22K18318, and No. 25K23360).

\bibliography{PHEC3}

\end{document}



\title{Supplemental Material \\
for ``Classical transport theory for the planar Hall effect with threefold symmetry''}

\author{Akiyoshi Yamada}
\author{Yuki Fuseya}%
\affiliation{%
Department of Physics, Kobe University, Kobe 657-8501, Japan
}

\date{\today}


\maketitle

\newpage

\renewcommand{\thefigure}{S\arabic{figure}}

\section{Three-ellipsoidal model}
We arrange three anisotropic ellipsoidal Fermi surfaces with threefold symmetry and obtain the Hall resistivity in the main text. 
Although the expression comprises many terms, extracting the coefficient proportional to $\sin 3\phi$ as a function of the field-rotation angle yields the following result:
\begin{eqnarray}
\rho_{xy}^{(3)}(B, \phi)&&=- \frac{B^3}{e n_e}  \frac{ (\mu_1 - \mu_2)^3 \cos^5\theta \sin\theta}{C_1 B^4 + C_2 B^2 + C_3}\sin3\phi
\end{eqnarray}
This term scales as $\sin\theta$ with the out-of-plane tilt angle $\theta$. 
This component remains finite when $\theta\neq0$.
In the high-field limit, the denominator is dominated by $B^{4}$, which causes this component to decay.

\section{Phase shift by sample rotation}
\begin{figure}[t]
\begin{center}
    \includegraphics[width=12cm]{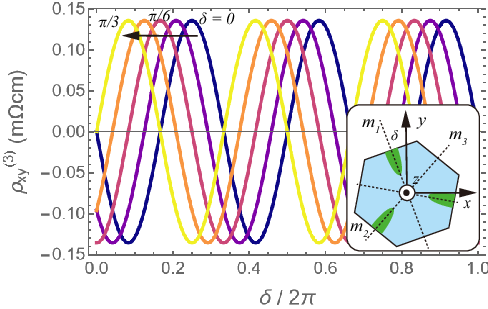}
\caption{\label{figS1_phase} Threefold component in PHE in the three-ellipsoidal Fermi surfaces and phase shift by rotating the surfaces. $m_{1}$, $m_{2}$, and $m_{3}$ indicate the mirror planes perpendicular to the $xy$ plane.}
\end{center}
\end{figure}

The three-fold component can include both $\sin3\phi$ and $\cos3\phi$ terms.
The weights of their coefficient can be changed continuously by rotating the $C_3$ symmetric Fermi surfaces, which yields a phase shift in the PHE signal.
We consider three anisotropic ellipsoidal Fermi surfaces as shown in the main text.
Each surface is tilted from the $xy$ plane by $20^\circ$ to break the mirror symmetry over this plane.
When we rotate the Fermi surfaces within the $xy$ plane by an angle $\delta$, the phase of the threefold component shifts [Fig.~\ref{figS1_phase}].
At $\delta=30^\circ$—where one mirror plane becomes parallel to the current~($x$)—the angular dependence reduces to a purely cosine form.
At $\delta = 60^\circ$, the angular dependence undergoes a complete sign inversion.

\section{Semimetals}
\begin{figure}[t]
\begin{center}
    \includegraphics[width=10cm]{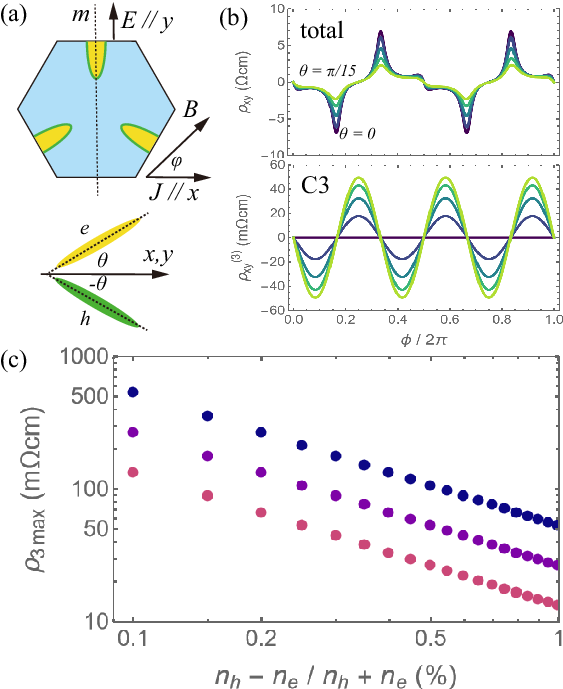}
\caption{\label{figS3_semimetal} (a) Schematic images for the six-ellipsoidal Fermi surfaces in a semimetal. (b) Field-angular dependence of PHE (upper panel) and the three-fold components (lower panel). $\mu_0=14$ T$^{-1}$, $\beta=0.005$, and residual hole density is $1\%$ of the total carrier number. (c) The maximum value of the threefold PHE amplitude as a function of the density of residual carrier.}
\end{center}
\end{figure}

We adopt the same threefold arrangement for ellipsoidal Fermi surfaces and assign opposite out-of-plane tilts $\theta$ to electron and hole pockets (Fig.~\ref{figS3_semimetal} (a)).
The results for the field angular dependence of PHE and the three-fold components are shown in Fig.~\ref{figS3_semimetal} (b).
The magnitude of the three-fold components exceeds that of a semiconductor with the same mobility.
This drastic enhancement originates from the charge compensation. Figure~\ref{figS3_semimetal} (c) shows the field- and angle-optimized maximum of the threefold component as a function of the density of residual carriers (holes). 
The amplitude increases as the residual carrier density decreases and the system approaches perfect compensation. 
Consequently, in semimetals, cleaner samples with fewer impurities are expected to display a more pronounced threefold component in PHE.

